\documentclass[aps,prl,twocolumn,superscriptaddress]{revtex4-1}
\usepackage{graphicx}
\usepackage{dcolumn}
\usepackage{bm}
\usepackage{color}
\usepackage{ulem}
\usepackage[usenames,dvipsnames,svgnames,table]{xcolor}

\newcommand{\Nd}{La$_{1.6-x}$Nd$_{0.4}$Sr$_x$CuO$_4$}

\newcommand{\pstar}{$p^{\star}$}
\newcommand{\Tstar}{$T^{\star}$}

\newcommand{\kzero}{$\kappa_0/T$}
\newcommand{\kN}{$\kappa_{\rm N}/T$}

\newcommand{\RH}{$R_{\rm H}$}
\newcommand{\nH}{$n_{\rm H}$}

\newcommand{\Tc}{$T_{\rm c}$}
\newcommand{\Hc}{$H_{\rm c2}$}

\begin{document}



\title{Wiedemann-Franz law and abrupt change in conductivity across the pseudogap critical point of a cuprate superconductor}

\author{B.~Michon}
\affiliation{Institut Quantique, D\'epartement de physique \& RQMP, Universit\'e de Sherbrooke, Sherbrooke, Qu\'ebec, Canada J1K 2R1}
\affiliation{Univ.~Grenoble Alpes, Institut N\'eel, F-38042 Grenoble, France}

\author{A.~Ataei}
\affiliation{Institut Quantique, D\'epartement de physique \& RQMP, Universit\'e de Sherbrooke, Sherbrooke, Qu\'ebec, Canada J1K 2R1}

\author{P.~Bourgeois-Hope}
\affiliation{Institut Quantique, D\'epartement de physique \& RQMP, Universit\'e de Sherbrooke, Sherbrooke, Qu\'ebec, Canada J1K 2R1}

\author{C.~Collignon}
\affiliation{Institut Quantique, D\'epartement de physique \& RQMP, Universit\'e de Sherbrooke, Sherbrooke, Qu\'ebec, Canada J1K 2R1}

\author{S.~Y.~Li}
\altaffiliation{Present address: State Key Laboratory of Surface Physics, Department of Physics, and Laboratory of Advanced Materials, Fudan University, Shanghai 200433, China}
\affiliation{Institut Quantique, D\'epartement de physique \& RQMP, Universit\'e de Sherbrooke, Sherbrooke, Qu\'ebec, Canada J1K 2R1}

\author{S.~Badoux}
\affiliation{Institut Quantique, D\'epartement de physique \& RQMP, Universit\'e de Sherbrooke, Sherbrooke, Qu\'ebec, Canada J1K 2R1}

\author{A.~Gourgout}
\affiliation{Institut Quantique, D\'epartement de physique \& RQMP, Universit\'e de Sherbrooke, Sherbrooke, Qu\'ebec, Canada J1K 2R1}

\author{F.~Lalibert\'e}
\affiliation{Institut Quantique, D\'epartement de physique \& RQMP, Universit\'e de Sherbrooke, Sherbrooke, Qu\'ebec, Canada J1K 2R1}

\author{J.-S.~Zhou}
\affiliation{Texas Materials Institute, University of Texas, Austin, USA}

\author{Nicolas~Doiron-Leyraud}
\email[]{nicolas.doiron-leyraud@usherbrooke.ca}
\affiliation{Institut Quantique, D\'epartement de physique \& RQMP, Universit\'e de Sherbrooke, Sherbrooke, Qu\'ebec, Canada J1K 2R1}

\author{Louis~Taillefer}
\email[]{louis.taillefer@usherbrooke.ca}
\affiliation{Institut Quantique, D\'epartement de physique \& RQMP, Universit\'e de Sherbrooke, Sherbrooke, Qu\'ebec, Canada J1K 2R1}
\affiliation{Canadian Institute for Advanced Research, Toronto, Ontario, Canada M5G 1Z8}

\date{\today}

\begin{abstract}

The thermal conductivity $\kappa$ of the cuprate superconductor \Nd~was measured down to 50~mK in seven crystals with doping from $p=0.12$ to $p=0.24$, 
both in the superconducting state and in the magnetic field-induced normal state.
We obtain the electronic residual linear term \kzero~as $T \to 0$ across the pseudogap critical point \pstar~$= 0.23$.
In the normal state, we observe an abrupt drop in $\kappa_0/T$ upon crossing below \pstar, consistent with a drop in carrier density $n$ from $1 + p$ to $p$, 
the signature of the pseudogap phase inferred from the Hall coefficient.
A similar drop in \kzero~is observed at $H=0$, showing that the pseudogap critical point and its signatures are unaffected by the magnetic field.
In the normal state, the Wiedemann-Franz law, \kzero~$= L_0 / \rho(0)$, is obeyed at all dopings, including at the critical point where the electrical resistivity $\rho(T)$ is $T$-linear down to $T \to 0$.
We conclude that the non-superconducting ground state of the pseudogap phase at $T=0$ is a metal whose fermionic excitations carry heat and charge as conventional electrons do.

\end{abstract}


\maketitle



\begin{figure}[t]
\includegraphics[scale=0.88]{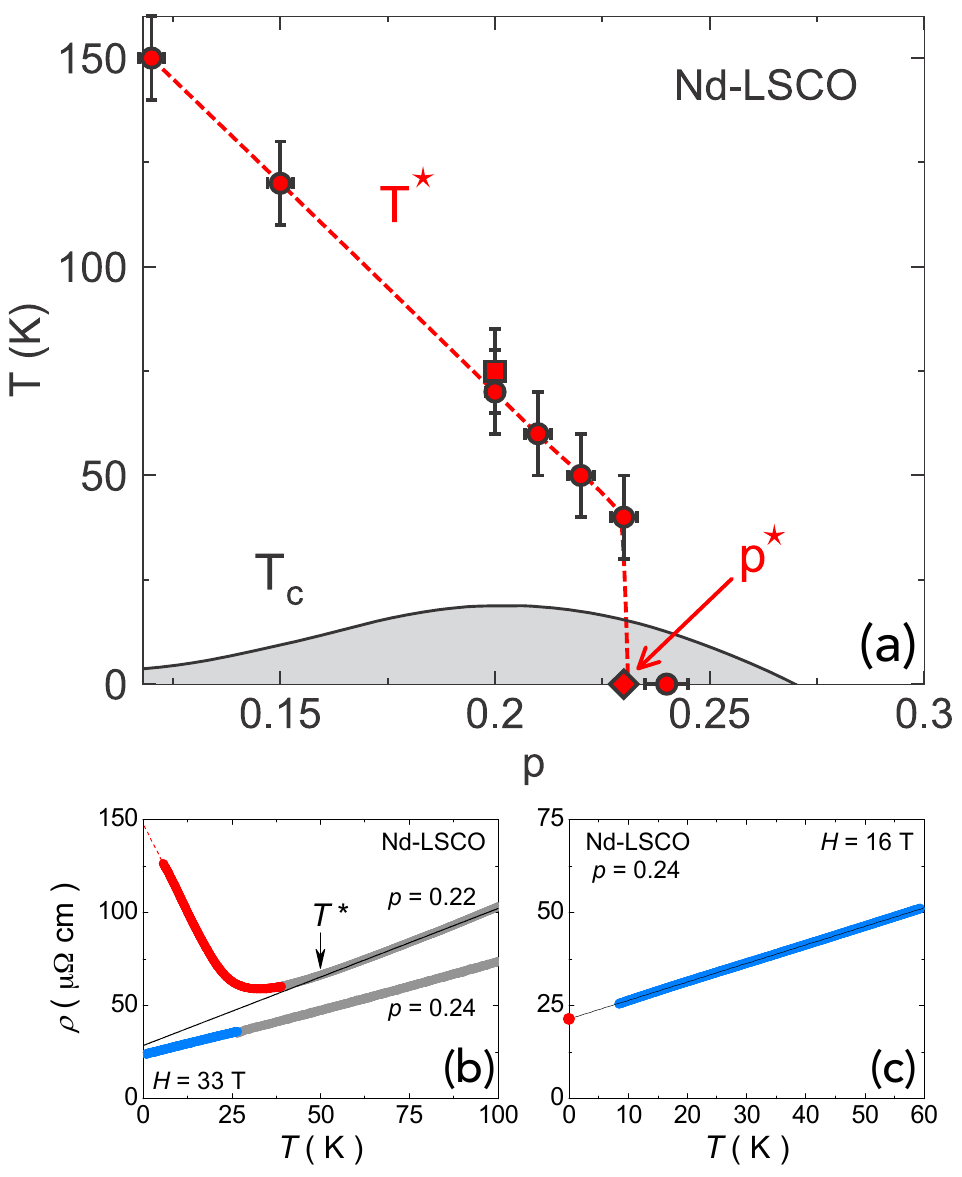}
\caption{
{\bf (a)}
Temperature-doping phase diagram of Nd-LSCO, showing the superconducting \Tc~(grey dome) 
and the pseudogap temperature \Tstar~extracted from the electrical resistivity (red dots; ref.~\cite{Collignon2017})
and from ARPES (red square; ref.~\cite{Matt2015}). 
The red diamond marks the position of \pstar~$= 0.23$, the doping for the onset of the pseudogap phase in Nd-LSCO.
The red dashed line is a guide to the eye.
{\bf (b)}
Electrical resistivity vs temperature for Nd-LSCO at $p$ = 0.22 and 0.24, at $H = 0$ (grey data) and in the normal state at $H = 33$~T (colored). 
The pseudogap temperature \Tstar~(arrow) is defined as the temperature below which $\rho(T)$ deviates from its $T$-linear behaviour at high temperature (black line). 
Here, \Tstar~$=50$~K at $p = 0.22$, and \Tstar~$=0$ at $p = 0.24$.
{\bf (c)}
Electrical resistivity of Nd-LSCO at $p$ = 0.24 and $H$ = 16~T (blue) with a linear fit (black line). The red dot is $L_0 / (\kappa_0/T)$, with $\kappa_0/T$ measured in the same sample at $H = 15$~T (Fig.~\ref{Fig2}),
showing that the Wiedemann-Franz law is perfectly satisfied.}
\label{Fig1}
\end{figure}

{\it Introduction. --}
Cuprate high-temperature superconductors exhibit a variety of correlated phases that interact with each other and with superconductivity, and understanding their associated complex phase diagram is a central challenge of condensed matter physics~\cite{Keimer2015}. The chief mystery is the pseudogap phase~\cite{Norman2005}, a phase that appears to break a number of symmetries, such as time-reversal~\cite{Fauque2006,Mook2008} and four-fold rotation~\cite{Daou2010,Sato2017}, below a temperature \Tstar, but whose fundamental nature is still unclear.
Several questions pertain to the critical doping \pstar~at which the pseudogap phase ends at $T=0$~\cite{Broun2008}.
At \pstar, the electrical resistivity remains $T$-linear as $T \to 0$~\cite{Cooper2009,Daou2009} (Fig.~\ref{Fig1}).
Does this imply a breakdown of the quasiparticle picture for the charge carriers?
Upon crossing below \pstar, the Hall number \nH~measured in the normal state of YBa$_2$Cu$_3$O$_y$ (YBCO),
reached by applying a large magnetic field,
is seen to drop dramatically~\cite{Badoux2016}, 
showing that the Fermi surface undergoes a rapid transformation upon entering the pseudogap phase.
The drop in \nH~has been attributed to a drop in carrier density $n$, from $n = 1 + p$ above \pstar~to $n = p$ below,
and explained in terms of a state that breaks translational symmetry~\cite{Storey2016,Eberlein2016,Verret2017,Chatterjee2017},
or not~\cite{Storey2016,Chatterjee2016,Morice2017}.
Alternatively, the drop in \nH~has been attributed to a nematic deformation of the Fermi surface~\cite{Maharaj2017}.

Below \pstar, the electrical resistivity $\rho(T)$ measured in the normal state of La$_{2-x}$Sr$_x$CuO$_4$ (LSCO) down to low temperature, reached by applying a large magnetic field, increases dramatically as $T \to 0$~\cite{Boebinger1996}. Originally interpreted in terms of a metal-to-insulator crossover upon cooling, the low-$T$ upturn in $\rho(T)$ has recently been attributed to a loss of carrier density below \Tstar~\cite{Laliberte2016}.
Is the upturn in $\rho(T)$ the result of localization or loss of carriers?
Are these various properties of charge transport measured in the presence of large magnetic fields the faithful signatures of the pseudogap phase unaltered by the field? Is the field a significant perturbation of the normal state itself?

Here we address these questions with measurements of heat transport in \Nd~(Nd-LSCO), a single-layer cuprate superconductor with a low critical temperature \Tc~and critical field \Hc, such that superconductivity can readily be suppressed with static fields down to $T \to 0$. 
In Fig.~\ref{Fig1}(a), the pseudogap phase of Nd-LSCO is delineated by its temperature \Tstar, defined as the temperature below which $\rho(T)$ departs from its $T$-linear behaviour at high temperature (Fig.~\ref{Fig1}(b)), in agreement with spectroscopic measurements of the pseudogap~\cite{Matt2015}. 
It ends at \pstar~$= 0.23 \pm 0.01$. At $p = 0.24$, $\rho(T)$ is seen to remain $T$-linear down to $T \to 0$ (Fig.~\ref{Fig1}(b)).
Hall measurements in Nd-LSCO find that \nH~$\simeq 1+p$ at $p = 0.24 >$~\pstar~and \nH~$\simeq p$ at $p = 0.20 <$~\pstar~\cite{Collignon2017}, in good agreement with YBCO~\cite{Badoux2016}.

The thermal conductivity $\kappa$ of Nd-LSCO was measured down to 50~mK in seven crystals, with $p$ ranging from 0.12 to 0.24 (see Table~\ref{T1}).
In summary, we find that the Wiedemann-Franz (WF) law is satisfied in the $T=0$ limit in the normal state of Nd-LSCO at all dopings.
This shows, that well-defined quasiparticles exist even at \pstar~and the pseudogap phase is a metal whose fermionic quasiparticles carry heat and charge as conventional electrons do.
A large drop in the electronic thermal conductivity \kzero~is observed upon crossing below \pstar, consistent with a drop of carrier density from $n \simeq 1+p$ to $n \simeq p$. Because a very similar decrease is seen in zero field, we conclude that the field does not affect the pseudogap phase or its transport signatures (other than by suppressing superconductivity).

%
%
%
%
%
%
\begin{figure}[t]
\includegraphics[scale=0.74]{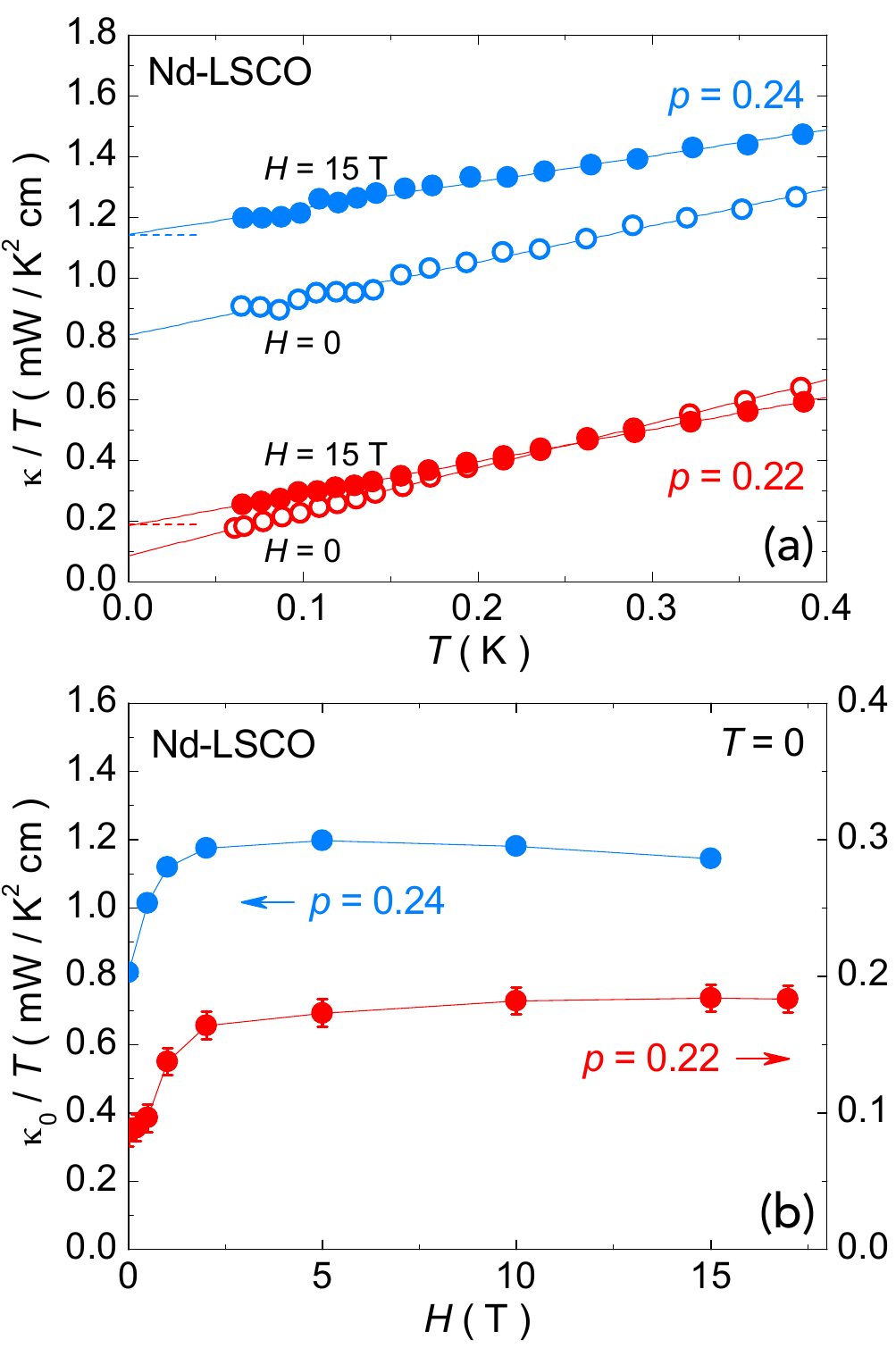}
\caption{
{\bf (a)}
Thermal conductivity $\kappa$ versus temperature plotted as $\kappa/T$ vs $T$, 
for Nd-LSCO at $p$ = 0.22 (red) and 0.24 (blue), in $H$ = 0 (open symbols) and 15~T (dots). 
The lines are linear fits to the data over the temperature range shown. 
The $y$-intercepts of the fits are the residual electronic terms $\kappa_0/T$. 
The horizontal dashed lines are calculated from the Wiedemann-Franz law L$_0/\rho$(0) using the measured $\rho$(0) (see text).
{\bf (b)}
$\kappa_0/T$ as a function of applied magnetic field for $p = 0.22$ (red) and 0.24 (blue). 
At both dopings, \kzero~saturates at high field, showing that the normal state has been reached. The error bars reflect the uncertainty on the fits shown in panel {\bf (a)}, which comes from varying the temperature range. For $p$ = 0.24 the error bars are smaller than the symbols.}
\label{Fig2}
\end{figure}
%
%

{\it Methods. --}
The thermal conductivity was measured on the same five single crystals of Nd-LSCO used in our previous study of electrical transport \cite{Collignon2017}, with $p = 0.20$, 0.21, 0.22, 0.23, and 0.24.
Details of the sample and contact preparation can be found there.
In addition, similarly prepared samples with $p = 0.12$ and 0.15 were measured. The \Tc~values for all samples are listed in Table~I.
The thermal conductivity was measured in the field-cooled state in a dilution refrigerator over the range 50~mK to 1.0~K, using a one-heater-two-thermometers steady-state technique. 
The heat current was applied in the $ab$ plane of the low-temperature tetragonal structure of Nd-LSCO and the magnetic field was applied along the $c$ axis.
%


{\it Results.--}
In Fig.~\ref{Fig2}(a), we show the thermal conductivity of Nd-LSCO for $p$ = 0.22 and 0.24 at $H = 0$ and 15~T, 
plotted as $\kappa/T$ vs $T$. 
As shown by the linear fits, the data below 0.4~K are well described by $\kappa/T = \kappa_0/T + BT$, where $\kappa_0/T$ is the electronic term and $B T$ is the phonon term. 
(The phonon conductivity $\kappa_{\rm ph}$ goes as $\kappa_{\rm ph} \sim T^{\alpha}$, with $\alpha = 2$ at high doping where the system is a good metal
and phonons are mainly scattered by electrons, as in overdoped Tl2201 \cite{Hawthorn2007},
and $\alpha > 2$ at low doping where the system is much more resistive.
The parameter $B$ is larger at $H=0$ because the density of quasiparticles that scatter phonons is lower
in the superconducting state.)
In Fig.~\ref{Fig2}(b), we plot \kzero~vs $H$ for both samples, showing how the conductivity increases with field from the superconducting state at $H=0$
until the normal state, reached at $H \simeq 10$~T for $p = 0.24$ and $H \simeq 15$~T for $p = 0.22$.
(Data at all fields are shown in SM Figs.~S1 and S2.)
In Fig.~3, the normal-state thermal conductivity at $H = 15$~T is displayed for all seven samples, with fits to extract \kzero.
(Data at $H$ = 0 are shown in SM Figs.~S3 and S4.)

\begin{table}[t]
\begin{tabular}{ccccccc}
\hline
$p$ & & \Tc &  \kzero  & $\rho(0)$ & $L_0$/$\rho(0)$ & $\rho_0$  \\
       & & (K) &  (mW/K$^2$cm)  & ($\mu\Omega$ cm) & (mW/K$^2$cm) & ($\mu\Omega$~cm) \\
\hline \hline
0.12 & & 5.0 &  0.036 & 600  & 0.041 & -- \\
0.15 & & 14.5 &  0.045 & 445  & 0.055 & -- \\

0.20 & & 15.5 &  0.105  & 229 & 0.106 & 46 \\
0.21 & & 15.0 &  0.083  & 253 & 0.096 & 59 \\
0.22 & & 14.7 &  0.184 & 138 & 0.177  & 29 \\
0.23 & & 12.4 &  0.410 & 60 & 0.410  & 43 \\
0.24 & & 10.7 &  1.144 & 21.4 & 1.140 & 21.4  \\
\hline
\end{tabular}
\caption{Doping $p$, superconducting \Tc, 
residual electronic term $\kappa_0/T$ at $H$ = 15 T (Fig.~\ref{Fig3}), 
normal state resistivity $\rho$(0) as $T \rightarrow$ 0 at $H$ = 15 T (see text), 
ratio $L_0$/$\rho$(0), and residual resistivity $\rho_0$ (see text) for all our measured Nd-LSCO samples. 
For $p = 0.24$, the values are at $H = 16$~T, 
except \Tc~which is in zero field. 
The uncertainty on \kzero~comes for the fits (Fig.~\ref{Fig3}) and is $\pm~0.01$~mW/K$^2$cm for all samples. 
The error bar on $\rho(0)$ comes from the extrapolation to $T = 0$ and $H = 15$~T, and is estimated to be $\pm~5~\mu\Omega$~cm. 
The uncertainty on $L_0$/$\rho(0)$ is calculated based on this error. 
The uncertainty on $\rho_0$ comes from the high temperature linear-$T$ fits (Fig.~\ref{Fig1}) 
and is $\pm~2~\mu\Omega$~cm. 
For $p = 0.24$, the error on $\rho(0)=\rho_0$ is $\pm~0.5~\mu\Omega$~cm, owing to the extended linear-$T$ regime down to low temperature.
}
\label{T1}
\end{table}

{\it Wiedemann-Franz law. --}
At $p = 0.24$, we make a precise test of the WF law, given by :
\[ \frac{\kappa_0}{T} = \frac{L_0}{\rho(0)} \]
where $\rho(0)$ is the electrical resistivity as $T \to 0$ and $L_0$ is the Sommerfeld value of the Lorenz number, 
equal to $2.44 \times 10^{-8}$~W$\Omega$/K$^2$. 
In Fig.~\ref{Fig1}(c), we plot $\rho$ vs $T$ measured in our Nd-LSCO sample with $p = 0.24$ at $H = 16$~T \cite{Collignon2017},
using the same contacts as for our $\kappa$ measurements.
The data are perfectly linear in temperature below $\sim 60$~K, down to $\sim 10$~K, the temperature below which 
$\rho(T)$ drops because of paraconductivity. 
(In Fig.~\ref{Fig1}(b), we see that applying 33~T confirms that $\rho(T)$ does remain linear down to at least 1~K \cite{Daou2009}.)
A linear extrapolation of the 16~T data yields $\rho(0)=21.4~\mu \Omega$~cm,
and therefore $L_0/\rho(0)=1.14$~mW/K$^2$cm.
This matches precisely the measured thermal conductivity, which yields
\kzero~$=1.14$~mW/K$^2$cm at $H = 16$~T, 
as indicated by the red dot at $T$ = 0 in Fig.~\ref{Fig1}(c).


\begin{figure}[t]
\includegraphics[scale=0.80]{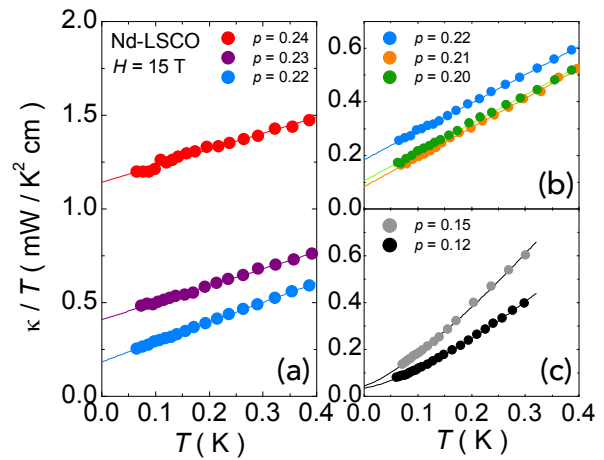}
\caption{
$\kappa/T$ versus temperature for Nd-LSCO at $H = 15$~T, 
for {\bf (a)} $p$ = 0.22, 0.23, and 0.24, {\bf (b)} $p$ = 0.20, 0.21, and 0.22, and {\bf (c)} $p$ = 0.12 and 0.15. 
In {\bf (a)} and {\bf (b)}, the lines are linear fits to the data over the entire range shown. 
In {\bf (c)} the lines are power-law fits to the data over the entire range of the data.
}
\label{Fig3}
\end{figure}


This shows that the WF law -- a fundamental property of conventional metals and Fermi liquids -- 
is precisely verified at the cuprate pseudogap critical point \pstar,
despite the fact that 
the resistivity exhibits the classic signature
of non-Fermi-liquid behaviour \cite{Lohneysen2007}, namely $\rho \propto T$ as $T \to 0$.
Moreover, the electronic specific heat of Nd-LSCO at $p = 0.24$ was recently shown to
exhibit the classic $T$ dependence associated with quantum criticality, namely 
$C_{\rm el} \propto - T$log$T$ as $T \to 0$ \cite{Michon2018}.
When combined, the three properties ($\rho$, $C_{\rm el}$ and $\kappa$) impose clear constraints on the nature of the pseudogap critical point.

The WF law was also tested in our six other samples, and found to hold in all cases, within error bars.
The values of \kzero~obtained from fits to the $H$ = 15~T data in Fig.~\ref{Fig3} are listed in Table~I.
We also list the values of $\rho(0)$ measured on the same samples with the same contacts, extrapolated to $T=0$ and to $H = 15$~T (data from ref.~\cite{Collignon2017} and SM Fig.~S5).
For example, in Fig.~\ref{Fig1}(b) the data for our $p = 0.22$ sample extrapolate to $147 \pm 5~\mu \Omega$~cm at $T = 0$ and $H = 33$~T.
Accounting for the magnetoresistance measured in that sample~\cite{Collignon2017}, we obtain $\rho(0) = 138 \pm 5~\mu \Omega$~cm at $H = 15$~T, and therefore $L_0 / \rho(0) = 0.177 \pm 0.010$~mW/K$^2$cm, which closely matches the measured \kzero~$=0.184 \pm 0.010$~mW/K$^2$cm at $H = 15$~T (Table I). The WF law is nicely satisfied.


\begin{figure}[t]
\includegraphics[scale=0.80]{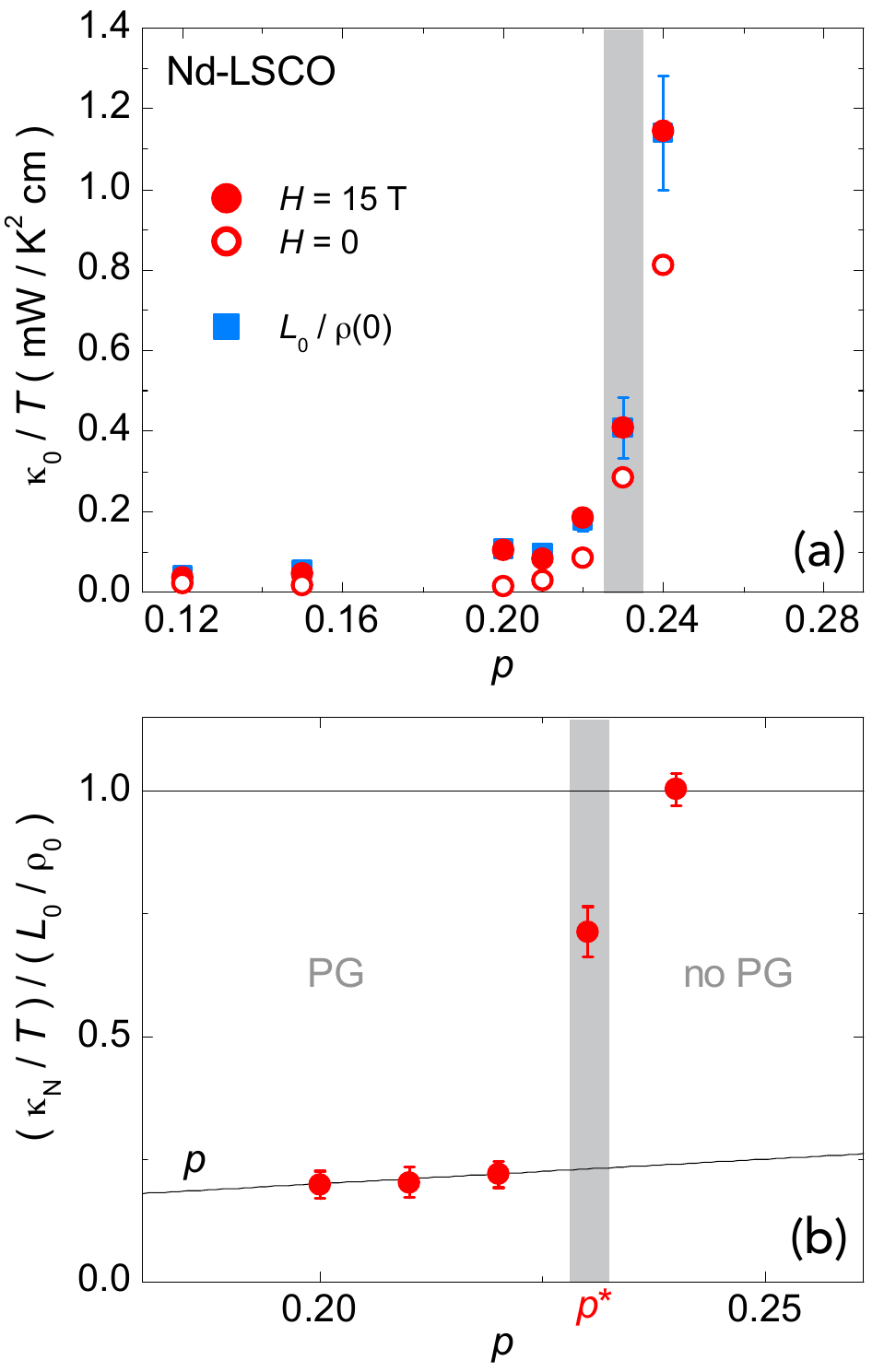}
\caption{
{\bf (a)}
$\kappa_0/T$ in Nd-LSCO at $H = 0$ (open red circles) and 15~T (red dots), and $L_0$/$\rho$(0) at $H = 15$~T (blue squares), as a function of doping. 
The error bars on $L_0$/$\rho$(0) come from the geometric factor error, $\pm$ 10\%, 
and the uncertainty on estimating $\rho$(0), $\pm~5~\mu\Omega$~cm ($\pm~0.5~\mu\Omega$~cm for $p = 0.24$). 
The error bar on \kzero~is $\pm~0.01$~mW/K$^2$cm, which is smaller than the symbols. 
{\bf (b)}
Ratio $(\kappa_{\rm N}/T)/(L_0/\rho_0)$ as a function of doping, where $\kappa_{\rm N}/T$ is the normal state $\kappa_0/T$, measured at $H = 15$~T, 
and $\rho_0$ is proportional to the level of disorder in each sample (see text).
The error bars come from the uncertainty on $\rho_0$, 
estimated to be $\pm~2~\mu\Omega$~cm ($\pm~0.5~\mu\Omega$~cm for $p$ = 0.24), 
and the error on \kzero~as in panel (a).
The grey vertical band in both panels gives the position of \pstar.
}
\label{Fig4}
\end{figure}


In Fig.~\ref{Fig4}(a), we plot \kzero~(red dots) and $L_0 / \rho(0)$ (blue squares) vs doping, both at $H = 15$~T,
for all 7 samples.
We find that the WF law is satisfied with $5 \%$ precision in the pure pseudogap phase, free of charge-density-wave (CDW) order
(in the doping interval between $p \simeq 0.18$ and \pstar),
as shown by our data at $p = 0.20$, 0.21, 0.22 and 0.23. 
This shows that the ground state of the enigmatic pseudogap phase (without superconductivity),
whatever its Fermi surface (closed pockets or arcs) and broken symmetries, 
has well-defined mobile fermionic excitations that carry heat and charge just as normal electrons do.

From our data at $p = 0.12$ and 0.15 (Fig.~\ref{Fig4}(a) and Table I), the WF law is also satisfied inside the CDW phase of Nd-LSCO ($0.08 < p < 0.18$),
as previously reported for the CDW phase of YBCO (in the transverse Hall channel, at $p = 0.11$) \cite{Grissonnanche2016}.
The WF law was also found to hold well above \pstar, in two strongly overdoped cuprates:
in Tl-2201 at $p = 0.3$,
where $\rho(T) = \rho_0 + A_1 T + A_2 T^2$, with $1 \%$ precision \cite{Proust2002},
and in LSCO at $p = 0.33$,
where $\rho(T) = \rho_0 + A_2 T^2$, with $20 \%$ precision \cite{Nakamae2003}.

{\it Drop in conductivity below \pstar. --}
With decreasing temperature, at fixed doping ($p <$~\pstar), the onset of the pseudogap phase at the crossover temperature \Tstar~(Fig.~\ref{Fig1}(a))
causes a large upturn in $\rho(T)$ at low $T$ (Fig.~\ref{Fig1}(b)).
In the $T=0$ limit, the fact that \kzero~is not zero but finite for $p <$~\pstar~and that it obeys the WF law, shows that
the ground state of the pseudogap phase is a metal and not an insulator.
Therefore, with decreasing $p$ at $T=0$, in the absence of superconductivity,
the transition that occurs at \pstar~is a metal-to-metal transition, and not a metal-to-insulator crossover.

We say `transition' because it is sharp.
This can be seen in Fig.~\ref{Fig4}(a), where the normal-state conductivity drops precipitously between $p = 0.24$ and $p = 0.21$ --
whether it is the electrical conductivity ($L_0 / \rho(0)$; blue squares) or the thermal conductivity (\kzero; red dots).
However, to be more precise in plotting the doping evolution of the conductivity, we need to factor out variations
in the level of disorder from sample to sample.

Hydrostatic pressure was recently shown to suppress the pseudogap and resistivity upturn in Nd-LSCO close to \pstar, revealing a linear-$T$ resistivity down to $T = 0$ at $p$ = 0.22 and 0.23 under 2~GPa~\cite{Doiron-Leyraud2017}.
This provides a direct measure of the intrinsic residual resistivity $\rho_0$ in the absence of the pseudogap phase.
An equivalent way to extract $\rho_0$ is to extrapolate linearly to $T=0$ the zero-field $T$-linear resistivity $\rho(T)$ above \Tstar, as shown by the solid line in Fig.~\ref{Fig1}(b), which yields the correct (pressure-revealed) $\rho_0$ for that sample.
(Note that disorder in cuprates is well-known to simply cause a rigid shift of the $T$-linear resistivity \cite{Rullier-Albenque2008}.)
In Table I, we list the value of $\rho_0$ thus obtained in all samples. We see that $\rho_0$ varies by a factor 3 or so.
In particular, the disorder level in our $p = 0.21$ sample is twice as large as in our $p = 0.22$ sample.
This is why the doping dependence of \kzero~is non-monotonic, with a local minimum at $p = 0.21$ (Fig.~\ref{Fig4}(a)).

In Fig.~\ref{Fig4}(b), we correct for the variation in disorder level by dividing \kzero~at $H=15$~T (red dots in Fig.~\ref{Fig4}(a)) by $L_0 / \rho_0$.
We now see that the conductivity evolves smoothly (and weakly) from $p = 0.22$ to $p = 0.20$.
(Note that because the same contacts are used to measure $\rho_0$ and \kzero~on each sample, there is no uncertainty
associated with geometric factors in the ratio (\kzero)$/ (L_0 / \rho_0)$.)
Given that the thermal conductivity normalized for disorder is 1.0 at $p = 0.24$, since $\rho(0) = \rho_0$ at that doping,
the drop down to the plateau at $p = 0.20-0.22$ occurs very rapidly, in an interval $\delta p \simeq 0.01$ at \pstar~$= 0.23$ (Fig.~\ref{Fig4}(b)).
This sharp drop reveals that the onset of the pseudogap phase at $T=0$ is a transition as a function of doping,
although it appears to be a crossover as a function of temperature.

It is interesting to examine the magnitude of this rapid drop in the $T=0$ conductivity across \pstar.
The normalized conductivity, (\kzero)$/ (L_0 / \rho_0)$, goes from 1.0 at $p = 0.24$ down to a value given by $p$ for $p$ = 0.22 and below.
This shows that in the pure pseudogap phase, the $T=0$ conductivity is a fraction $p$ of its full value 
in the absence of the pseudogap, when the metal has its large  Fermi surface.
This large and sudden drop in conductivity at \pstar, by a factor $\sim 5$, is a clear signature of the pseudogap transition.
It is consistent with the drop in carrier density $n$ inferred from the Hall effect in YBCO \cite{Badoux2016}
and Nd-LSCO \cite{Collignon2017}, from $n \simeq 1+p$ at $p>$~\pstar~to $n \simeq p$ at $p<$~\pstar.
Specifically, in the same samples of Nd-LSCO, the Hall number \nH~drops by a factor 5 between $p = 0.24$ and 
$p = 0.20$, where \nH~$\simeq p$, and so does the conductivity.

Model calculations of transport properties across a quantum phase transition where AF order sets in \cite{Storey2016,Chatterjee2017,Verret2017}
are able to reproduce the drop in \nH~seen in YBCO \cite{Badoux2016} and Nd-LSCO \cite{Collignon2017}, as
expected from the Luttinger rule given the reconstruction of the Fermi surface imposed by the AF Brillouin zone.
However, the calculated change in the associated conductivity (at $T=0$) is smaller than what we observe in Nd-LSCO, 
roughly by a factor 2 \cite{Chatterjee2017}.
The pseudogap phase seems to have this interesting property that the conductivity suffers the full loss of carrier density,
as already noted for LSCO \cite{Laliberte2016}.
This large drop in conductivity is difficult to explain in a scenario of nematic order \cite{Nie2014}, 
for such order does not reduce the carrier density, it only changes the Fermi surface shape and curvature \cite{Maharaj2017}.

{\it Superconducting state} ($H = 0$).--
Turning to the zero-field data (Fig.~\ref{Fig2} and SM Figs. S3 and S4), we observe a finite and sizable residual electronic thermal conductivity in the superconducting state (Fig.~\ref{Fig4}(a)).
This is due to transport by $d$-wave nodal quasiparticles. 
In the clean limit where the impurity scattering rate $\Gamma_0$ is much smaller than the $d$-wave gap maximum $\Delta_0$,
\kzero~is `universal', {\it i.e.}, independent of $\Gamma_0$, and only dependent on the quasiparticle velocities $v_{\rm F}$ and $v_{\Delta}$ \cite{Graf1996,Durst2000,Shakeripour2009}.
As $\Gamma_0$ increases and \kN~$\sim 1 / \Gamma_0$~decreases, \kzero~increases and eventually becomes a sizable fraction of \kN,
when $\Gamma_0$ becomes comparable to $\Delta_0$ \cite{Sun1995}. 
In that dirty limit, \kzero~mimicks \kN.
This is the limit we are in with all our Nd-LSCO samples.
For our $p=0.24$ sample, with $\rho_0 = 21.4~\mu \Omega$~cm and \Tc~$= 10.7$~K (Table I), we estimate that 
$\Gamma_0 \simeq \Delta_0$.
In that sample, 
\kzero~$= 0.81$~mW/K$^2$ cm at $H=0$ (Fig.~\ref{Fig2}), which is 70\% of the normal-state value, \kN, measured at $H=15$~T (Fig.~\ref{Fig2}).
Note that even the significantly cleaner crystals of overdoped Tl2201, with $\rho_0 \simeq 6~\mu \Omega$~cm and \Tc~$= 15$~K,
are in the dirty limit, with \kzero~$\simeq 0.3$~\kN~\cite{Proust2002}.
Note also that such high normal-state fractions, due to strong pair breaking by disorder, necessarily imply low superfluid densities,
perhaps as low as $\simeq 10$~\% of the carrier density \cite{Sun1995}, as found in samples of overdoped LSCO with values of $\rho_0$ and \Tc~comparable to our Nd-LSCO samples~\cite{Bozovic2016}.

In Fig.~\ref{Fig4}(a), we see that \kzero~(at $H=0$) closely tracks \kN~(at $H=15$~T) as a function of doping. In particular, it exhibits a very similar drop below \pstar.
This allows us to draw two important conclusions.
First, the pseudogap critical point is present in the superconducting state, at a location unchanged from the normal state,
as found recently in Raman studies of Bi2212 \cite{Loret2017}.
Superconductivity does not seem to affect the pseudogap phase very much.
Secondly, the similarity between high-field and zero-field thermal conductivity data shows that the pseudogap signatures seen 
in the Hall coefficient and electrical resistivity are not high-field effects, 
but essentially zero-field phenomena.
In other words, the loss of carrier density deduced from transport measurements (of \RH, $\rho$ or $\kappa$) is independent of magnetic field.


{\it Summary. --}
We measured the thermal conductivity of Nd-LSCO across its pseudogap critical point \pstar~$=0.23$.
In the field-induced normal state, the fermionic conductivity \kzero~at $T=0$ drops precipitously when $p$ falls below \pstar,
in tandem with the drop in the Hall number \nH~\cite{Collignon2017}, 
confirming that the pseudogap phase is characterized by a drop in carrier density \cite{Badoux2016}.
At $H=0$, \kzero~exhibits a very similar drop below \pstar, showing that the drop in carrier density is not
a high-field effect and \pstar~is not shifted by the field.
The WF law is precisely satisfied at $p = 0.24$, even if the charge carriers exhibit non-Fermi-liquid behavior at that doping, 
namely a resistivity that remains $T$-linear down to the lowest temperatures \cite{Daou2009}.
The WF law is also satisfied at $p <$~\pstar, showing that the pseudogap phase has fermionic excitations
that conduct heat and charge as normal electrons do.
In the superconducting state, 
\kzero~at $H=0$ is $\simeq 70$~\% of its normal-state value, showing that there is strong pair breaking in our Nd-LSCO crystals.
This implies that the superfluid density must be very small, 
as indeed found in LSCO films with similar disorder levels \cite{Bozovic2016}.


{\it Acknowledgements. --}
L.T. acknowledges support from the Canadian Institute for Advanced Research (CIFAR) 
and funding from the Natural Sciences and Engineering Research Council of Canada (NSERC; PIN:123817), 
the Fonds de recherche du Qu\'ebec - Nature et Technologies (FRQNT), 
the Canada Foundation for Innovation (CFI), 
and a Canada Research Chair. 
This research was undertaken thanks in part to funding from the Canada First Research Excellence Fund. 
Part of this work was funded by the Gordon and Betty Moore FoundationÕs EPiQS Initiative (Grant GBMF5306 to L.T.).



%

\pagebreak

\onecolumngrid

\setcounter{figure}{0}
\renewcommand{\thefigure}{S\arabic{figure}}

\begin{center}
{\Large Supplementary Material for\\
``Wiedemann-Franz law and abrupt change in conductivity across the pseudogap\\
 critical point of a cuprate superconductor"}
\end{center}

%
%
\begin{figure}[b]
\includegraphics[scale=2.0]{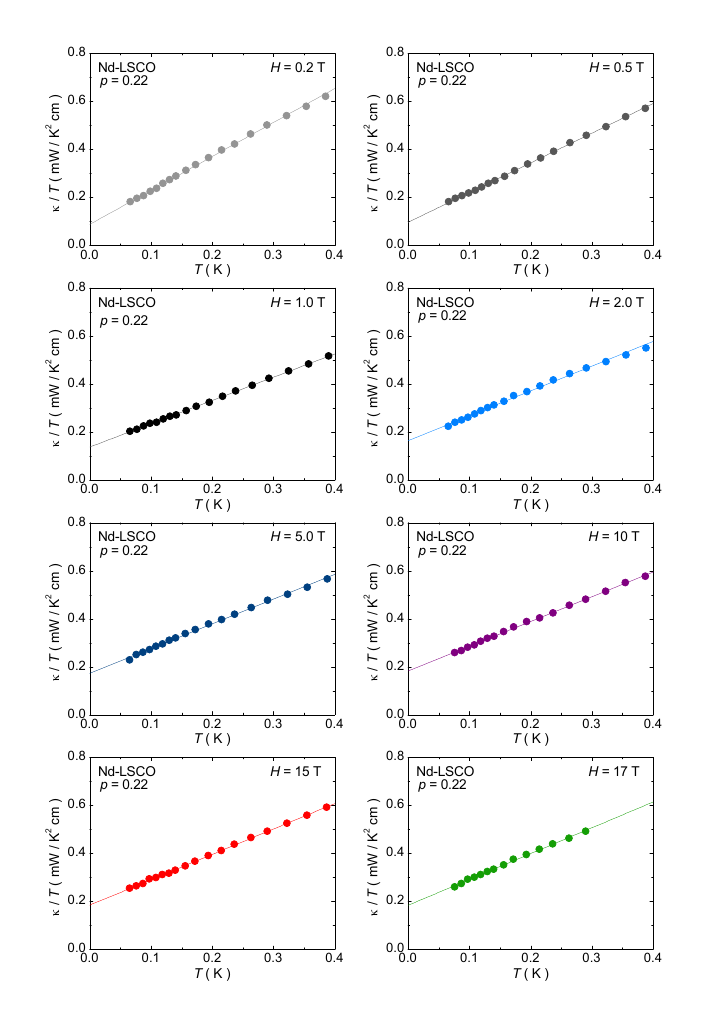}
\caption{
Thermal conductivity $\kappa$ versus temperature plotted as $\kappa/T$ vs $T$ for Nd-LSCO at $p$ = 0.22, in magnetic fields as indicated. In all panels the line is a linear fit to the data over the entire range shown.
}
\end{figure}
%
%
%
%
\begin{figure}[t]
\includegraphics[scale=2.0]{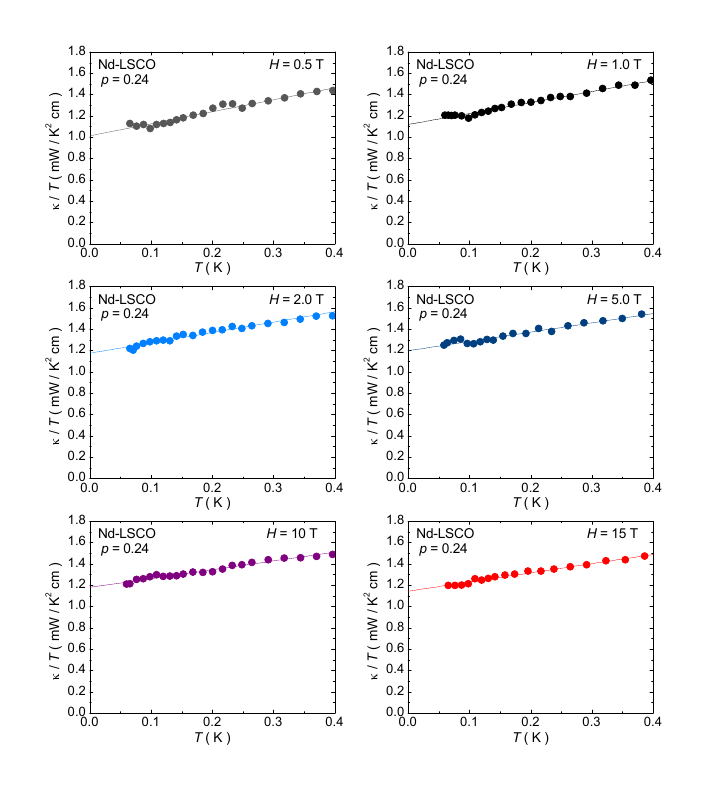}
\caption{
Thermal conductivity $\kappa$ versus temperature plotted as $\kappa/T$ vs $T$ for Nd-LSCO at $p$ = 0.24, in magnetic fields as indicated. In all panels the line is a linear fit to the data over the entire range shown.
}
\end{figure}
%
%


\begin{figure}[t]
\includegraphics[scale=2.0]{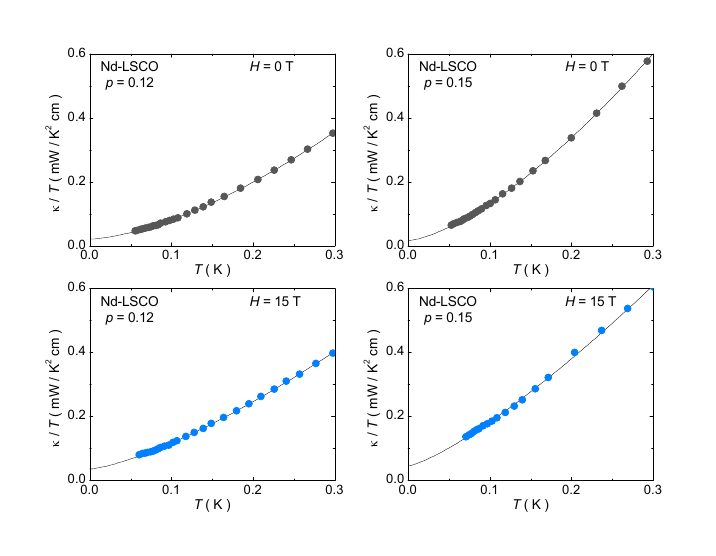}
\caption{
Thermal conductivity $\kappa$ versus temperature plotted as $\kappa/T$ vs $T$ for Nd-LSCO at $p$ = 0.12 and 0.15, in magnetic fields as indicated. In all panels the line is a power-law fit of the form $\kappa/T = a + bT^{\alpha}$ to the data over the entire range shown. For $p$ = 0.12 the values of $\alpha$ are 1.56 and 1.37 in $H$ = 0 and 15~T, respectively. For $p$ = 0.15 the values of $\alpha$ are 1.47 and 1.28 in $H$ = 0 and 15~T, respectively.
}
\end{figure}



\begin{figure}[t]
\includegraphics[scale=2.5]{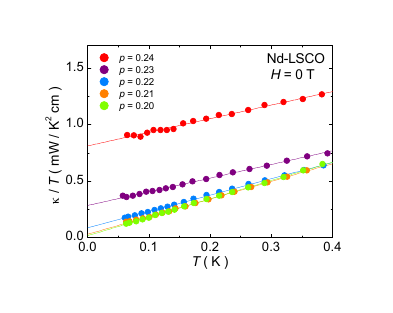}
\caption{
Thermal conductivity $\kappa$ versus temperature plotted as $\kappa/T$ vs $T$ for Nd-LSCO at dopings as indicated, in zero field. The lines are linear fits to the data over the entire range shown
}
\end{figure}



\begin{figure}[t]
\includegraphics[scale=2.0]{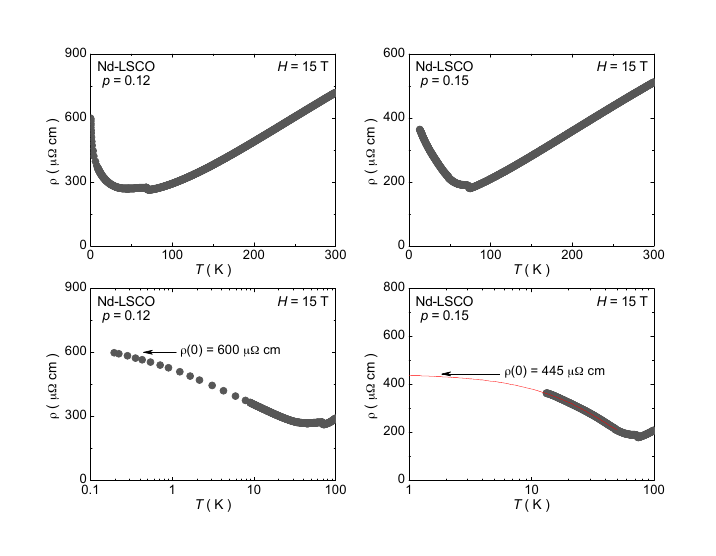}
\caption{
Electrical resistivity of Nd-LSCO at $p$ = 0.12 and 0.15 in $H$ = 15~T, shown on a linear scale up to 300~K (top), and on semi-log scale at low temperature below 100~K.
}
\end{figure}


\end{document}